\documentclass[aps,pra,twocolumn,superscriptaddress,floatfix]{revtex4-2}
\usepackage[utf8]{inputenc}
\usepackage[english]{babel}
\usepackage{amsmath}
\usepackage{amsfonts}
\usepackage{amssymb}
\usepackage[margin=2cm]{geometry}
\usepackage{graphicx}
\usepackage{bm}
\usepackage{hyperref}
\usepackage[dvipsnames]{xcolor}
\usepackage[normalem]{ulem}
\usepackage{natbib}
\addto\captionsenglish{}

\newcommand{\dxd}{\int d{\bf x}}
\newcommand{\DQ}{\int_{\mbf{Q_i}}^{\mbf{Q_F}}D\mbf{Q}}

\newcommand{\intk}{\int_{\mathcal{C}}}
\newcommand{\mbf}[1]{\mathbf{#1}}

\newcommand{\ifuc}{\Phi_{\mathcal{C}}[\mbf{Q}]}

\newcommand{\xy}{\mbf{x}-\mbf{y}}
\newcommand{\qiu}{\mbf{q}_{i}'}
\newcommand{\qid}{\mbf{q}_{i}''}
\newcommand{\qju}{\mbf{q}_{j}'}
\newcommand{\qjd}{\mbf{q}_{j}''}
\newcommand{\rg}{\mbf{r}}

\newcommand{\re}{\text{Re}}
\newcommand{\im}{\text{Im}}

\begin{document}
\author{Matteo Sighinolfi}
\affiliation{INO-CNR  BEC  Center, 38123, Trento, Italy } 
\affiliation{Dipartimento di  Fisica,  Universit\`a di Trento, 38123  Trento, Italy}
\author{Davide De Boni}
\affiliation{Dipartimento di  Fisica,  Universit\`a di Trento, 38123  Trento, Italy}
\author{Alessandro Roggero}
\affiliation{Dipartimento di  Fisica,  Universit\`a di Trento, 38123  Trento, Italy}
\affiliation{Trento Institute for Fundamental Physics and Applications, INFN, 38123, Trento, Italy}
\author{Giovanni Garberoglio}
 \affiliation{
   European Centre for Theoretical Studies in Nuclear Physics and
        Related Areas (FBK-ECT*), Trento, I-38123, Italy.}
        \affiliation{Trento Institute for Fundamental Physics and Applications, INFN, 38123, Trento, Italy}
\author{Pietro Faccioli}
\affiliation{Dipartimento di  Fisica,  Universit\`a di Trento, 38123  Trento, Italy}
\affiliation{Trento Institute for Fundamental Physics and Applications, INFN, 38123, Trento, Italy}\email{pietro.faccioli@unitn.it}
\author{Alessio Recati}
\affiliation{INO-CNR  BEC  Center, 38123, Trento, Italy } 
\affiliation{Dipartimento di  Fisica,  Universit\`a di Trento, 38123  Trento, Italy}
\affiliation{Trento Institute for Fundamental Physics and Applications, INFN, 38123, Trento, Italy}\email{alessio.recati@cnr.it}

\title{Stochastic Dynamics and Bound States of Heavy Impurities  in a Fermi Bath }
\date{\today}
\begin{abstract}
We investigate the dynamics of heavy  impurities embedded in an ultra-cold Fermi gas by using a Generalized Langevin equation. The latter -- derived by means of influence functional theory -- describes  the stochastic classical dynamics of the impurities and the quantum nature of the fermionic bath manifests in the emergent interaction between the impurities and in the viscosity tensor. By focusing on the two-impurity case, we predict the existence of bound states, in different conditions of coupling and temperature, and whose life-time can be analytically estimated. Our predictions should be testable using cold-gases platforms within current technology.  
\end{abstract}
\maketitle

\section{Introduction}

The concept of mediated interactions between particles due to the medium they are immersed in is ubiquitous in physics. 
Notable examples include the phonon-mediated interaction between electrons \cite{Mahan} --giving rise to Bardeen--Cooper--Schrieffer superconductivity--,  the interaction between cluster or nuclear pasta structures mediated by the surrounding neutron fluid in the inner crust of neutron stars \cite{BULGAC2001695,bulgac2001casimir,magierski2002neutron}, and the interaction between heavy quarks mediated by a plasma of deconfined quarks and gluons, in  super-hot hadronic matter \cite{ROTHKOPF20201}. 

Highly imbalanced mixtures of ultra-cold gases provide clean and tunable platforms where to study medium mediated interactions.  In these systems, the quasi-particles resulting from dressing  impurities by the polarization of the bath are usually referred to as  polarons. 
The study of polaron physics in cold gases was initiated by  seminal experimental works on the normal-to-superfluid phase transition in imbalanced Fermi-Fermi mixtures \cite{Shin2006,Partridge503} and the identification of the normal phase as a weakly interacting gas of polarons, in the spirit of Landau Fermi liquid theory (see, e.g., \cite{Shin2008,Lobo2008} and reference therein). 
Shortly after, also the case of impurities immersed in a Bose gas was experimentally realized \cite{Jin2016,arlt2016}.   

Presently, the static and dynamical properties of a single polaron have been relatively well understood, at least for the case of a degenerate polarized Fermi bath, at zero temperature~\cite{massignan2014polarons}. On the other hand, the experimental  and theoretical characterization of the effect of the mediated interaction between impurities is, in general, much more challenging~\cite{camacho2018bipolarons}.
However,  two very recent experiments have measured the effect of the mediated interaction on a Bose condensed gas in a Bose-Fermi mixture, in which  the Fermi gas plays the role of the bath~\cite{DeSalvo2019,DavidsonMediated2020}

The present work aims at exploring the dynamics of heavy impurities in a Fermi bath at finite temperature, within the framework of Generalized Langevin equation (GLE) \cite{Lampo2017bosepolaronas} that is derived from a chain of well-controlled approximations, starting from the a microscopic Feynman--Vernon influence functional \cite{feynman2000theory}. In this way, we are able to provide semi-analytical expressions for both the mediated inter-impurity  interaction and the configuration-dependent friction tensor.   

Within our approach the effective stochastic dynamics of the haevy impurities is treated at the classical level. On the other hand, quantum effects must be fully taken into account when deriving the mediated interaction -- which corresponds to a finite temperature Ruderman–-Kittel-–Kasuya–-Yosida (RKKY)\cite{ruderman1954indirect, kasuya1956theory, yosida1957magnetic} potential -- and the friction tensor. 
 
As a case study, we analyse the dynamics of two heavy impurities. By numerically integrating their stochastic equations of motion starting from  configurations in which they are close to each other, we find evidence for the formation of a transient bound state. Numerical estimates of the life-time of this state at different temperatures  agree well with the analytic calculations of the dissociation rate performed within  Kramers' theory (see, e.g., \cite{hanggi-1990-reaction} and reference therein), thus demonstrating that the impurity pair dissociation is a thermally activated rare event.

We also find that the position-dependent off-diagonal elements in the friction tensor have important implications on the dynamics of the pair.  In particular, the relative motion of two close impurities is almost frictionless, yet the presence of a longitudinal and a transverse friction leads to a rapid dissipation of the relative  orbital angular momentum (see Fig.~\ref{Fig: orbit}~b)).

\begin{figure}[t]

\begin{minipage}[t]{0.5\textwidth}
       \includegraphics[keepaspectratio=true,width=8.6cm]{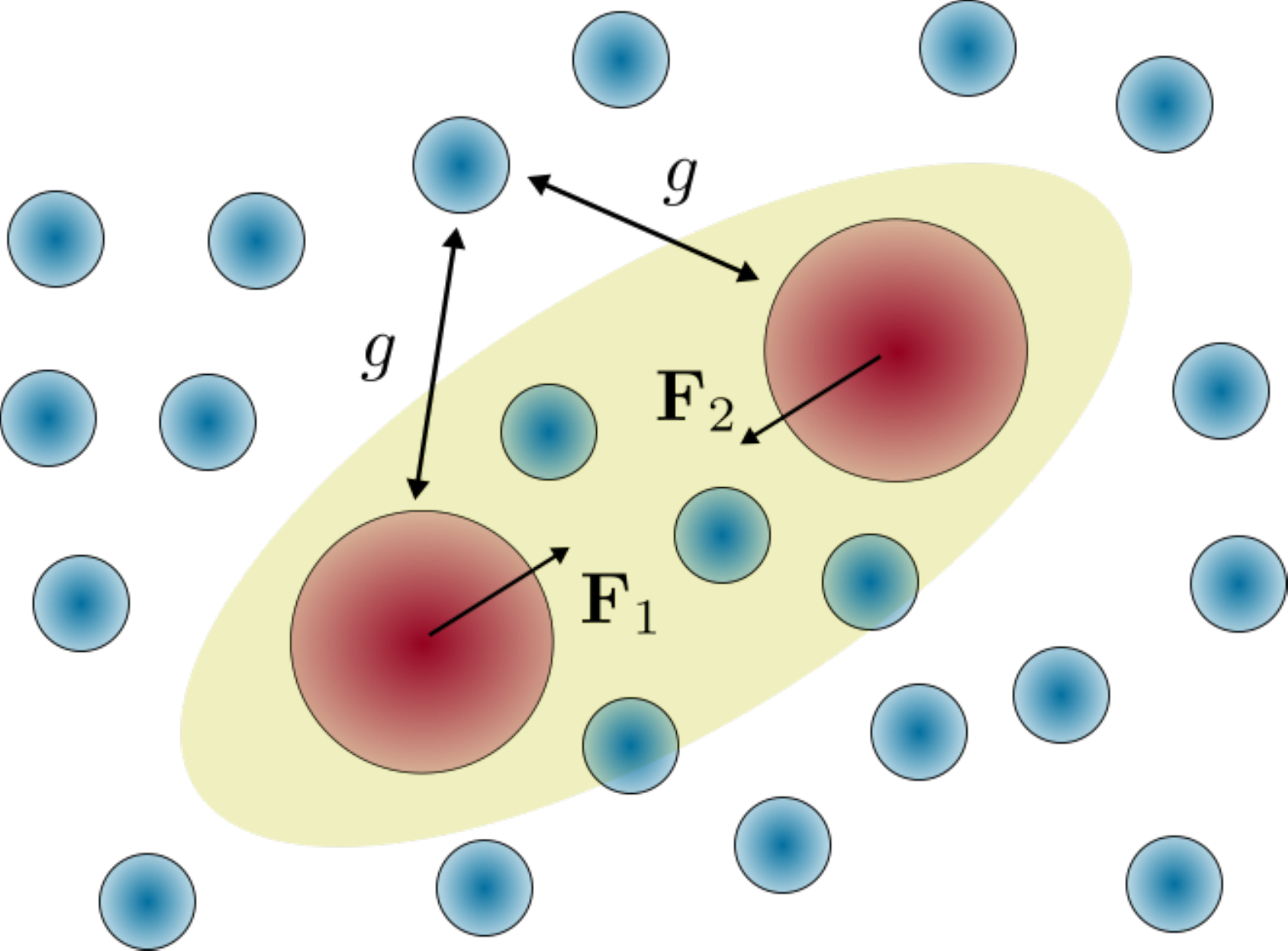}
       \end{minipage}
     
\caption{\label{Fig: 2 impurities} 
Two heavy impurities (red) in a bath of fermions (cyan): the bare impurity-bath interaction $g$ is responsible for the induced forces between impurities ${\bf F}_{1,2}$ and for the low friction region (yellow).
 }
\end{figure}

The paper is organized as follows: in Sec. \ref{Sec: model and theory}, we derive the Feynman--Vernon influence functional for our system; in Sec. \ref{Sec: stoch dynamics}, we describe the quantum mediated interaction and friction and derive the GLE for the dynamics of the impurities, focusing on the one- and two-impurity cases; in Sec. \ref{Sec: numerical results}, we discuss the numerical results obtained for the dynamics of two impurities which are initially close to each other. A summary of our findings is the content of Sec. \ref{Sec: conclusions}.

\section{Theoretical Setup} \label{Sec: model and theory}

We consider a system composed by a bath of degenerate ultra-cold Fermi atoms of mass $m$ and chemical potential $\mu$, interacting with $N$ impurities of mass $m_I\gg m$. 
At the  energy scales we consider, particles interact only via s-wave scattering and therefore the interaction between the atoms of the bath can be neglected. For the sake of clarity we also consider that there is no direct interaction between the impurities.
The interaction between the bath and the impurities is characterized in the following by a contact potential with strength $g$. 
We also assume that the system is at a temperature $T$ such that the de Broglie thermal wavelength of the impurities $\lambda\sim \hbar\sqrt{2\pi/m_I k_B T}$ is small compared to their typical inter-particle distance. This will allow us to regard impurities as quasi-classical particles. 
In order to obtain a stochastic equation of motion for the impurities, it is convenient to describe them in first quantization and coordinate representation. We rely on quantum field theory to describe the dynamics of the degenarate fermionic bath. Our system can be modelled by the following  Hamiltonian:
\begin{equation}
\hat H=\hat H_F+\hat H_{I}+\hat V \, , \label{total Hamiltonian}
\end{equation} 
where 
\begin{align}
& \hat H_{I}=\sum_{i=1}^N \frac{\hat{ \mathbf{p}}_i^2}{2m_I}\\
&\hat H_F=\dxd ~\hat \Psi^{\dagger}({\bf x })\left( \frac{-\hbar^2}{2m}\hat \nabla^2-\mu \right)\hat \Psi({\bf x}) \\
&\hat V=g \sum_{i=1}^N\dxd ~ \hat \Psi^{\dagger}(
{\bf x})~\delta(\hat {\bf q}_i-{\bf x})~\hat\Psi({\bf x}), 
\end{align}
where $\hat{\mbf{p}}_i$ and $\hat{\mbf{q}}_i$  denote the impurity momentum and position operators,  $\hat \Psi({\bf x})$ and $\hat\Psi^\dagger({\bf x})$ are the annihilation and creation field operators for the particles in the bath.

Let us  consider a setup in which the impurities are initially decoupled from the bath and localized at fixed positions ${\bf Q}_i\equiv (\mbf{q}_1,\ldots,\mbf{q}_N)$.
 At time $t=0$, the interaction with the bath is switched on and the system's  density matrix begins to evolve according to the  Hamiltonian~(\ref{total Hamiltonian}). We are interested in the diagonal elements of the reduced density matrix for the impurities, i.e. in the probability of observing the impurities at  ${\bf Q}_f=(\mbf{q}^f_1,\ldots,\mbf{q}^f_N)$ at time $t_f$. Using Feynman--Vernon path integral representation of the density matrix~\cite{feynman2000theory} we obtain:
\begin{equation}
P(\mbf{Q}_f,t|\mbf{Q}_i,0)=\int_{\mbf{Q}_i}^{\mbf{Q}_f} \mathcal{D}\mbf{Q}\int \mathcal{D}\xi \int \mathcal{D}\xi^{*}e^{\frac{i}{\hbar}S[\mbf{Q},\xi,\xi^{*}]}. \label{prob trans definition}
\end{equation}
In this equation,   $\xi(t,{\bf x})$ and $\xi^*(t,{\bf x})$ are Grassmann coherent field variables, while the functional at the exponent is
\begin{equation}
\begin{split}
S[\mbf{Q},\xi,\xi^{*}]&=\intk dt' 
\left\lbrace \frac{m_I}{2}\sum_{j=1}^N \dot{\mbf{q}}^2_j(t')+\right. \\
 & \left. \hspace{-2cm}\dxd \xi^{*}(t', {\bf x}) \left(i\hbar \partial_{t'} -\frac{\hbar^2\nabla^2}{2m}-\mu-\rho(t',{\bf x}) \right)\xi(t', {\bf x})\right\rbrace,
\end{split} \label{total action}
\end{equation}
where
$\rho(t, {\bf x})=g\sum_{i=1}^N\delta(\mbf{q}_i(t)- {\bf x})$ is the instantaneous impurity density  and 
the time integral is defined over the standard Keldysh contour $\mathcal{C}$~\cite{kamenev2011field,calzetta2008nonequilibrium}. 

The integral over the Grassmann fields $\xi, \xi^*$  can be carried out analytically, leading to: 
\begin{equation}
P(\mbf{Q}_f,t|\mbf{Q}_i,0)=\DQ~ e^{i\ifuc}~\\
e^{i\frac{m_I}{2\hbar}\sum_{j=1}^N \intk dt' \dot{\mbf{q}_j}^2}, \label{prob trans}
\end{equation}
where $\ifuc$ is the influence functional, which is formally written as 
\begin{equation}
i \ifuc =  \text{Tr} \left[\log \left(i\hbar \partial_{t'} - \frac{\hbar^2 \nabla^2}{2m}-\mu -\rho(t', {\bf x})  \right) \right]. \label{inf func general}
\end{equation}
To obtain an explicit representation for $\ifuc$, it is convenient to deal separately with the upper and lower branches of the Keldysh contour. In addition, we assume a low impurity density and  perform a functional expansion to second order in   $\rho(t', {\bf x})$. The $0$-th order term is a constant that is reabsorbed in the definition of probability, while the first order term is an energy shift that does not affect the dynamics.  
 The resulting expression for the transition probability density is 
 \begin{eqnarray}
    P(\mbf{Q}_f,t|\mbf{Q}_i,0) = \int_{{\bf Q}_i}^{\bf Q_f} \mathcal{D}{\bf Q'}\int_{{\bf Q}_i}^{\bf Q_f} \mathcal{D}{\bf Q''}\nonumber\\
    e^{i\Phi({\bf Q'}, {\bf Q''})}~e^{i\frac{m_I}{2\hbar}\sum_{j=1}^N 
    \int_{0}^t dt' \left( \dot{\mbf{q}}_j^{'2}-\dot{\mbf{q}_j}^{''2}
    \right)},
    \label{prob trans nokeldish}
\end{eqnarray}
where
\begin{eqnarray}
\Phi({\bf Q}', {\bf Q}'') = \frac{i }{2}\sum_{a,b=1}^2\int_0^t~dt' \int_0^t dt^{''} \int d{\bf x} \int d{\bf y}  \nonumber\\
\rho_a(t^{'},{\bf x})
\Delta_{ab}(t^{'}-t^{''},{\bf x}-{\bf y})~\rho_b(t^{''},{\bf y}),
\label{inf func not expanded}
\end{eqnarray}
where $a$, $b$ label the branches of the Keldysh contour ${\cal C}$, primed variables lie on the forward branch of the contour and double-primed variables lie on the backward branch. In particular in Eq.~(\ref{inf func not expanded}), 
\begin{eqnarray}
\rho_1 (t, {\bf x})= g \sum_{i} \delta({\bf q}_i^{'}(t)- {\bf x})\nonumber\\
\rho_2 (t, {\bf x})= g \sum_{i} \delta({\bf q}_i^{''}(t)- {\bf x}),
\label{rhos}
\end{eqnarray}
and $\Delta_{ab}$ are the entries of a $2\times 2$ matrix of Green's functions: 
\begin{eqnarray}
\label{Delta F}
\Delta_{11}(t, {\bf x})&=&\Delta_F (t, {\bf x})= i D_F(t, {\bf x}) \nonumber\\
\Delta_{12}(t, {\bf x}) &=&-\Delta_<(t, {\bf x}) =-iD_<(t, {\bf x})  \nonumber\\ 
\Delta_{21}(t, {\bf x}) &=&-\Delta_>(t, {\bf x})=-iD_>(t, {\bf x})  \nonumber\\ \Delta_{22}(t, {\bf x}) &=& \Delta_{\tilde{F}}(t, {\bf x})=iD_{\tilde{F}}(t, {\bf x}).
\end{eqnarray}
Here,  $D_>(t,{\bf x}), D_<(t,{\bf x})$ and $D_F(t,{\bf x})$ are the standard fermionic polarization propagators of many-body theory~\cite{fetter2012quantum, giuliani2005quantum}.
 
We emphasize that the expressions~(\ref{prob trans nokeldish}) and~(\ref{inf func not expanded}) follow directly from Eq.~(\ref{prob trans}), in the small impurity density limit. 
 
In the next section, we shall introduce additional approximation which enable us to efficiently compute this transition probability by integrating stochastic differential equation of motions. 

\section{Effective Stochastic Dynamics of Heavy Impurities} \label{Sec: stoch dynamics}

In this section, we introduce a chain of well-controlled approximations to enable the sampling of the transition probability density~(\ref{prob trans}). 

\subsection{Small Frequency Expansion} 

Since the mass of the impurities is much greater than that of the particles in the bath the dynamics of the former is expected to be much slower.   
Then, it is possible to perform a small frequency expansion of $\Delta_{ab}$ in Eq.~(\ref{Delta F}):
\begin{eqnarray}\label{Delta taylor}
        \Delta_{ab}(t, {\bf x}) &=&
        \int \frac{d \omega}{2 \pi} e^{-i \omega t}\left(
        \sum_{n=0}^\infty \frac{\omega^n}{n!}~F_{ab}^{(n)} (x) \right)\nonumber\\
    &=&F^{(0)}_{ab}({\bf x} ) + i \frac{d}{dt} \delta(t) F^{(1)}_{ab}({\bf x})+ \ldots, \label{smalle freq expansion}
\end{eqnarray}
where 
\begin{eqnarray}
F^{(0)}_{ab}(\xy) &\equiv& \Delta_{ab}(\omega=0,\xy)\\
F^{(1)}_{ab}(\xy)&\equiv& \lim_{\omega \to 0}\frac{d}{d\omega}\Delta_{ab}(\omega,\xy),
\end{eqnarray}
and the dots denote higher order terms in the Taylor expansion.  

Substituting Eqs.~(\ref{Delta taylor},\ref{rhos}) into Eq.~(\ref{inf func not expanded}) we obtain
\begin{equation}
\begin{split}
\Phi({\bf Q}', {\bf Q}'') 
&=\frac{ig^2}{2}\sum_{i,j = 1}^N\int_0^t du\bigg\lbrace F^{(0)}_F(\qiu-\qju)+\\
+&F^{(0)}_{\tilde{F}}(\qid-\qjd)-F^{(0)}_<(\qiu-\qjd)+\\
-&F^{(0)}_>(\qid-\qju)-i\dot{\mbf{q}}_{j1}\frac{\partial}{\partial \qju}F_>^{(1)}(\qid-\qju)\\
-&i\dot{\mbf{q}}_{j2}\frac{\partial}{\partial \qjd}F_<^{(1)}(\qiu-\qjd) \bigg\rbrace. \label{inf func expanded}
\end{split} 
\end{equation}

It is convenient to introduce  the so-called complex potential $\mathcal{V}(\xy)$:
\begin{equation}
i\mathcal{V}(\xy) \equiv F^{(0)}_F(\xy)=V(\xy)+iW(\xy). \label{complex potential}
\end{equation}
 In appendix~\ref{Appendix A}, we  show that the real and imaginary part of  $\mathcal{V}$ can be expressed in terms of the retarded polarization propagator in Fourier space:
\begin{align}
&V(\xy)=\text{Re}D^R(\omega=0,\xy)  \label{real potential}\\ 
&W(\xy)=\frac{2}{\beta}\lim_{\omega \to 0}\frac{1}{\omega}\text{Im}D^R(\omega,\xy). \label{imaginary potential}
\end{align}
We also show that for a bath of non-interacting fermions in 3 dimensions
\begin{eqnarray}
V(\xy) = -\frac{m k_F}{4\pi^4\hbar^2}\int dq \frac{\sin(qr)}{r} \int dk \,  f_{FD}(k,T) k  \times \nonumber \\
\times \log \left\vert \frac{k+q/2}{k-q/2} \right\vert \nonumber \\ \label{real potential position} \\
  W(\xy) = -\frac{m^2}{2\pi^3\hbar^3\beta} \int dq \,  f_{FD}(q/2,T) q \frac{\sin(qr)}{qr} \nonumber, \\ \label{imaginary potential position}
\end{eqnarray}
where $r=|\xy|$ and $f_{FD}$ is the Fermi-Dirac distribution.

In the following, we use the rescaled imaginary potential $W_R$ as
\begin{equation}
W_R(\xy) = \frac{\beta}{2}W(\xy), \label{rescaled potential}
\end{equation}
for an easier understanding. Indeed, with this rescaling the term $1/\beta$ in Eq.~(\ref{imaginary potential position}) disappears.

\subsection{Classical Limit } 
We now take the classical limit for the dynamics of the impurities. In order to implement this approximation,  we first perform the change of variables
\begin{equation}
\mbf{r}_i=\frac{1}{2}(\qiu+\qid) \qquad \mbf{y}_i=\qiu-\qid. \label{semiclassical approx}
\end{equation}
After an integration by parts, the free action of the impurity takes the form: 
\begin{equation}
    e^{\frac{i m_I}{\hbar} \sum_{i=1}^N\int_0^t 
\ddot{\bf r}_i\cdot {\bf y}_i }.
\end{equation}
We expect the dominant contribution to the path integral to come from the functional region where the time integral in the exponent is small or at most of order unity. 
To estimate it, we note that
$\int_{t_i}^{t_f} dt \ddot{\bf r}_i \cdot {\bf y}_i\sim  \sqrt{k_B T/m_I} |{\bf y}_i |$, 
where $\sqrt{k_B T/m_I}$ is the average thermal velocity of the impurities. 
Then, the stationary phase condition implies $|{\bf y}_i | \lesssim \sqrt{1/ m_I k_B T}.$
In the limit of heavy impurities, fluctuations of  $y(t)$ become small compared to all relevant length scales, thus we can expand the influence functional to second order in ${\bf y}_i$, leading to
\begin{equation}
\begin{split}
&P(\mbf{R}_f,t | \mbf{R}_i,0)=\int_{\mbf{R}_i}^{\mbf{R}_f}{\cal D}\mbf{R}\int_0^0 \mathcal{D}\mbf{Y} \\
&\text{exp}\left\lbrace -\frac{i}{\hbar}\int_0^t dt' \left[\mbf{y}_i \left( m_I\ddot{\mbf{r}}_i+\Gamma_{ij}(\mbf{R})\dot{\mbf{r}}_j-\mbf{F}_i(\mbf{R}) \right)+ \right.\right. \\
& \left.\left. -\frac{1}{2}\mbf{y}_i \frac{2}{\beta} \Gamma_{ij}(\mbf{R})\mbf{y}_j\right] \right\rbrace, 
\end{split} \label{trans probab final}
\end{equation}
where ${\bf R}=({\bf r}_1, \ldots, {\bf r}_N)^T$, ${\bf Y}=({\bf y}_1, \ldots, {\bf y}_N)^T$ and the sum over repeated indices $i,j = 1,\ldots , N$ is understood. $F(\mbf{R})$ and $\Gamma_{ij}(\mbf{R})$ are defined as 
\begin{align}
&\mbf{F}_i(\mbf{R})=-g^2 \sum_{j=i}^N \nabla V(\mbf{r}_i-\mbf{r}_j) \label{forces} \\
&\Gamma_{ij}(\mbf{R})=g^2 \mathcal{H}_{W_R}(\mbf{r}_i-\mbf{r}_j), \label{hessian}
\end{align}
where $\mathcal{H}_{W_R}$ is the Hessian of $W_R$.

The Gaussian integral over ${\bf Y}$ can be evaluated analytically, leading to our final path integral expression for the transition probability: 
\begin{eqnarray}
P(\mbf{R}_f,t | \mbf{R}_i,0)=\int_{\mbf{R}_i}^{\mbf{R}_f}{\cal D} \mbf{R}
e^{-\int_0^t d\tau \left(m_I \ddot {\bf R} - m_I \Gamma({\bf R}) \dot {\bf R} - {\bf F}({\bf R})\right)^2}.\nonumber\\
\label{prob trans rf ri}
\end{eqnarray}
Here, the probability for the impurities to go from ${\bf R}_i$ to ${\bf R}_f$ in a time $t$ is written as a functional integral over all possible trajectories connecting the initial and the final configuration. 
 We note that the functional  at the exponent, which determines the relative statistical weight of ${\bf R}(t)$ trajectories, does not  explicitly depend on $\hbar$. 
 Indeed, it corresponds to an Onsager--Machlup action~\cite{OnsMach}, which characterizes path integral representation of propagator in classical Fokker--Planck dynamics.   

As a consequence, as explicitly shown in Ref. \cite{blaizot2016heavy,lau2007state},  the same transition  probability density of Eq.~(\ref{prob trans rf ri}) can be generated by the following GLE: 
\begin{equation}
m_I\ddot{\mbf{r}}_i=-\Gamma_{ij}(\mbf{R})\dot{\mbf{r}}_j+\mbf{F}_i(\mbf{R})+\mbf{\Psi}_i(\mbf{R},t). \label{generalized langevin equation (complete)}
\end{equation}
The  viscosity  $\Gamma(\mbf{R})_{ij}$ and the noise term  $\mbf{\Psi}_i(\mbf{R},t)$ satisfy the fluctuation-dissipation relations
\begin{align}
&\langle \mbf{\Psi}_i(\mbf{R},t) \rangle=0 \label{+noise single expect value complete} \\
&\langle \mbf{\Psi}_i(\mbf{R},t)\otimes\mbf{\Psi}_j(\mbf{R},t')\rangle=\frac{2}{\beta} \Gamma_{ij}(\mbf{R})\delta (t-t'). \label{noise double expect value complete}
\end{align}
The noise $\mbf{\Psi}_i{(\mbf{R})}$ depends only the relative distances between the $i$-th impurity and all the others impurities.
To conclude this section, we note that while the  dynamics of the impurities has been reduced to a classical diffusion process, the quantum nature of the bath is still effectively encoded in the structure of the viscosity and force terms, derived from Eqs.~(\ref{real potential position}, \ref{imaginary potential position}) \footnote{Within our approximations, the impurity mass $m_I$ is not renormalized. However, this effect can be considered performing the small frequency expansion at order $\omega^2$, where terms proportional to $\ddot{{\bf r}}_i$ and to $\dot{{\bf r}}_i^2$ are present. Moreover, in the limit of $m_I \gg m$ and in the range of interaction considered is known to be small, see e.g.~\cite{combescot2007normal}. } .

\begin{figure*}[t]

\begin{minipage}[t]{0.48\textwidth}
         \includegraphics[keepaspectratio=true,width=8.6cm]{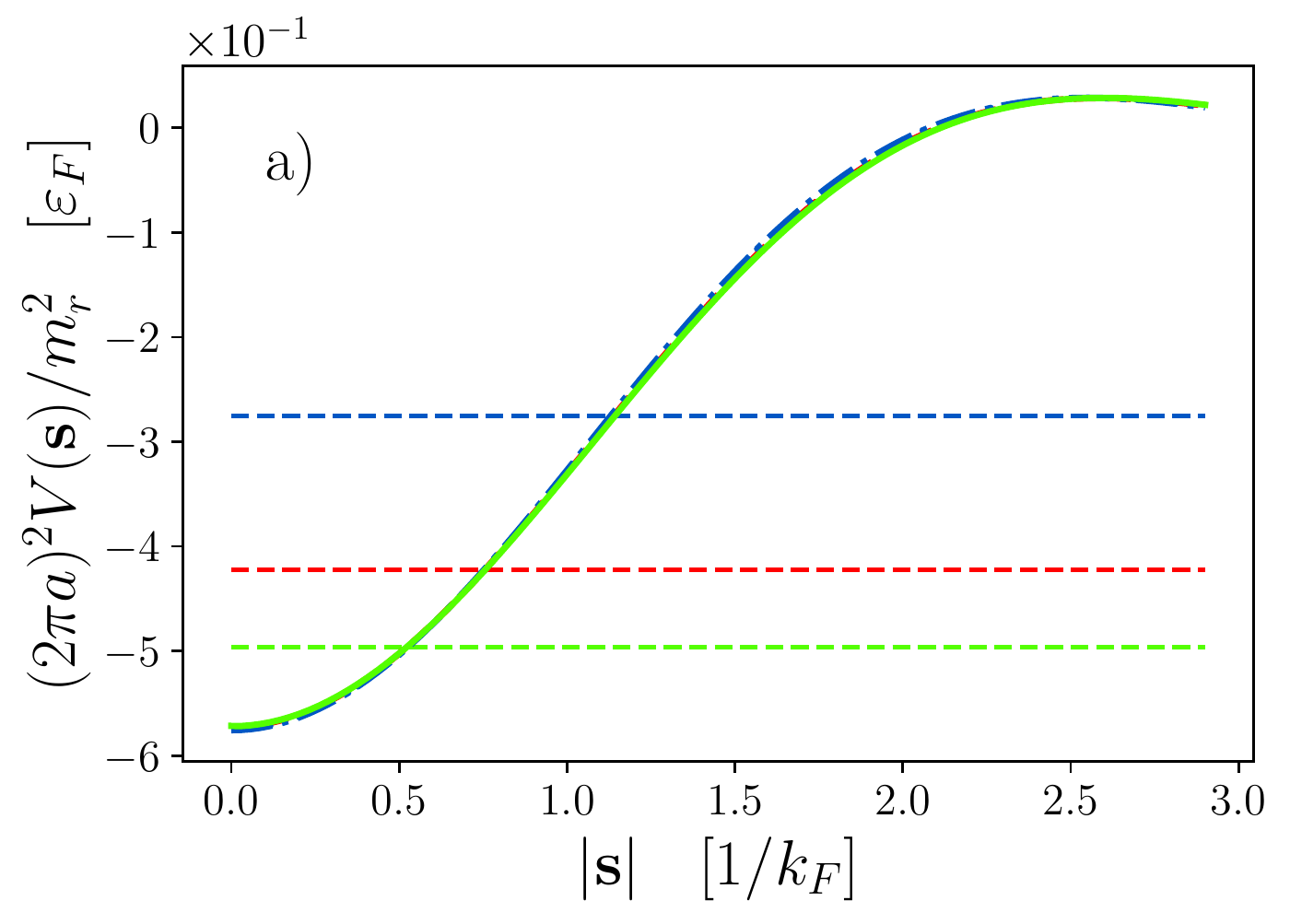}

         \label{Fig: V and Gamma a}
     \end{minipage}
     \hfill
     \begin{minipage}[t]{0.48\textwidth}
        \includegraphics[keepaspectratio=true,width=8.6cm]{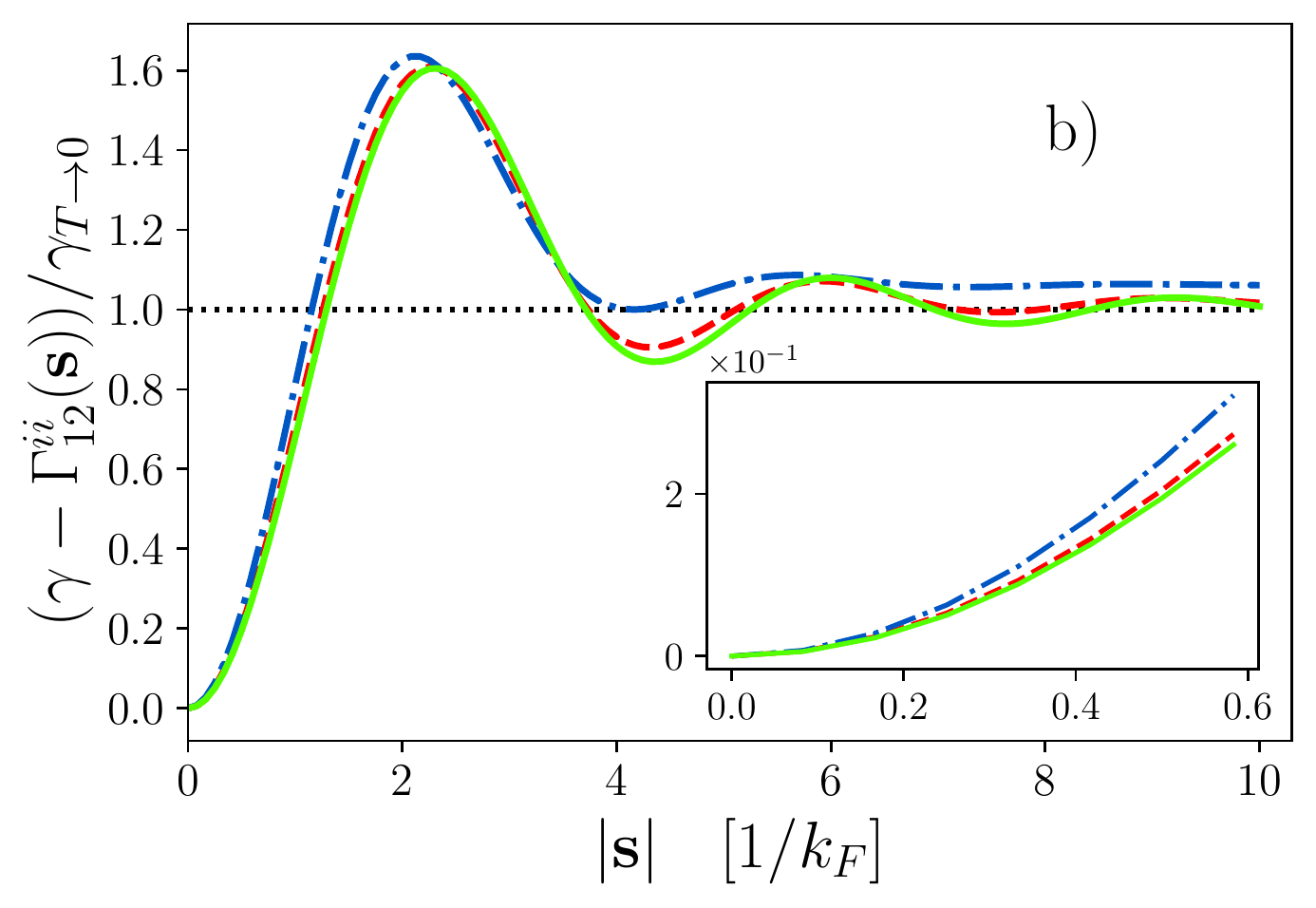}
     \end{minipage}
     \hfill
     \begin{minipage}[t]{0.48\textwidth}
        \includegraphics[keepaspectratio=true,width=8.6cm]{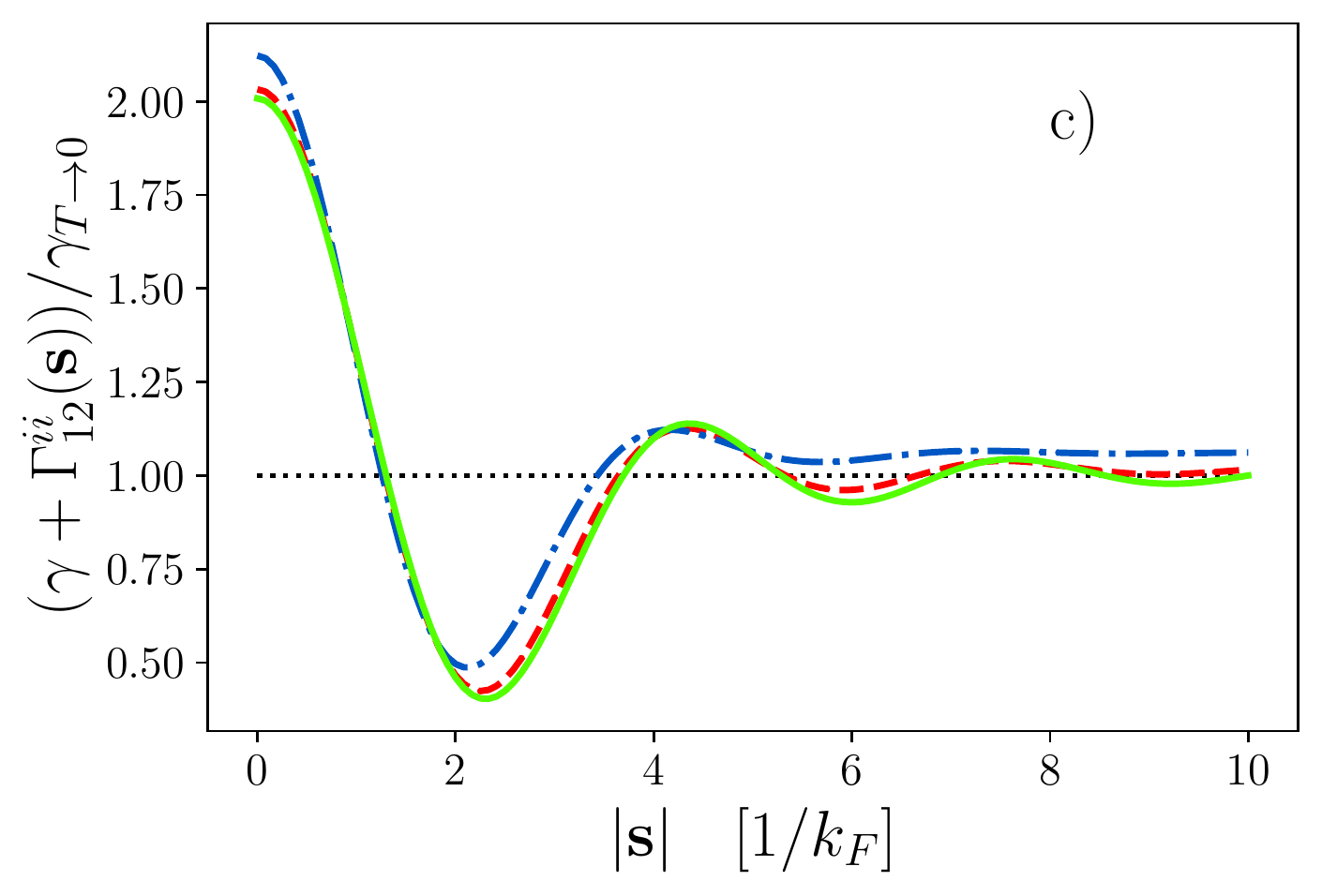}
     \end{minipage}
     \hfill
     \begin{minipage}[t]{0.48\textwidth}
        \includegraphics[keepaspectratio=true,width=8.6cm]{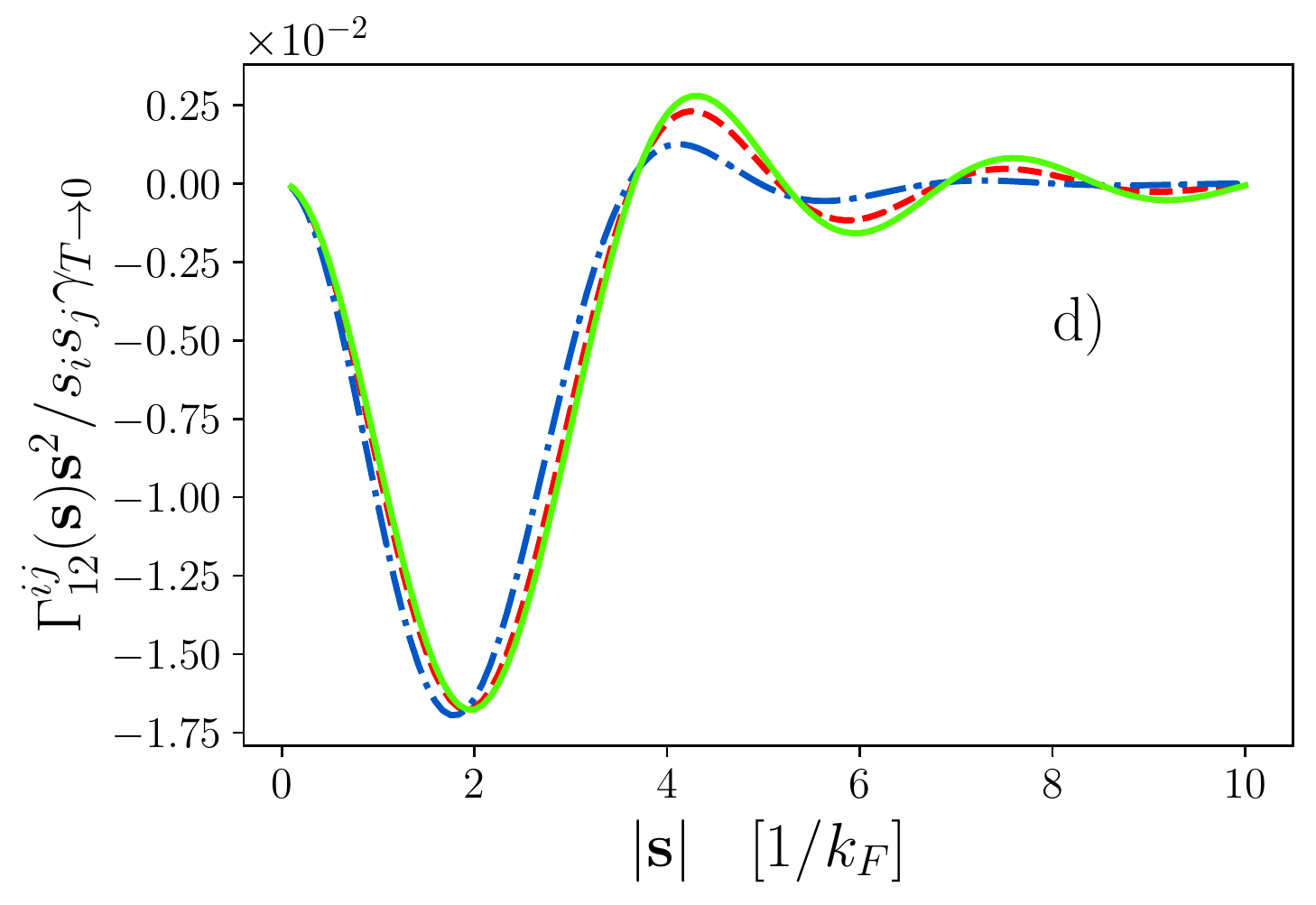}
     \end{minipage}
\caption{\label{Fig: V and Gamma} Some spatial dependences with $k_F a =0.19$ and $T/T_F =0.2$, $0.1$, $0.05$ (dotted-dashed blue, dashed red, full green lines). Panel a): $V({\bf s})$ at $k_F a =0.19$, the intersection between horizontal dashed lines and potential curves determines the typical size of the bound state $r_ {\textrm b}$. Panel b): difference between the constant viscosity term and the diagonal component of the viscosity matrix $\gamma-\Gamma_{12}^{ii} ({\bf s})$ in units of $\gamma_{T \to 0}$. Note that $|{\bf s}| k_F \gtrsim 6$ the oscillations decay because the contribution of $\Gamma_{12}^{ii}$ vanishes. In the inset is shown the behaviour for small $|{\bf s}|$. Panel c): sum between the constant viscosity term and the diagonal component of the viscosity matrix $\gamma-\Gamma_{12}^{ii} ({\bf s})$ in units of $\gamma_{T \to 0}$. Panel d): off-diagonal component of the viscosity matrix $\Gamma_{12}^{ij}$ multiplied by $|{\bf s}|^2/s_i s_j$ in units of $\gamma_{T \to 0}$. Also for the off-diagonal term the oscillatory behaviour decays for $|{\bf s}| k_F \gtrsim 6$.}
\end{figure*}

\subsection{Dynamics of a single impurity} \label{Subsec: generalized Langevin equation for 1 impurities}

It is instructive to first apply our formalism to the case of a single impurity. For $N=1$, the GLE reduces to that of a standard Brownian particle, with constant viscosity and white noise:
\begin{equation}\label{Lang1}
m_I \ddot{\mbf{r}}=-\gamma\dot{\mbf{r}}+\mbf{\Psi}(t), 
\end{equation}
where we defined $\mbf{r}=\mbf{R}=\mbf{r}_1$ and 
\begin{equation}\label{viscT>0}
\gamma= \Gamma_{11}=-\frac{8 m^4 g^2}{3 \hbar^7 \pi^3}(k_B T)^2\text{Li}_2(-e^{\beta \mu(T)}),  
\end{equation}
is the single impurity friction constant with Li$_2$ is the dilogarithm \cite{gradshteyn2014table}.
At finite temperature, Eqs. (\ref{Lang1}) and (\ref{viscT>0}) yield the conventional Einstein's diffusion law, and the kinetic energy of the impurity thermalizes with the bath. However, if temperature is much smaller then the bath Fermi temperature $T_F=\varepsilon_F / k_B$, with the usual Fermi energy $\varepsilon_F = \hbar^2 k_F^2/2m$, Eq.(\ref{viscT>0}) can be written as
\begin{equation}
    \gamma_{T \to 0}=\frac{4 \hbar k_F^2}{3 \pi}\left(\frac{m}{m_r} k_F a \right)^2\left(1+\frac{T^2}{T_F^2}\frac{\pi^2}{3} \right). \label{gamma T to 0}
\end{equation}
In this equation,  $m_r=m_I m/(m_I+m)$ is the reduced mass and $g$ is expressed in terms of the more physical $s$-wave scattering length $a$, $g=2\pi\hbar^2 a/m_r$.  We note that the viscosity remains finite even at zero temperature. This is possible because  the impurity releases energy into the bath, by inducing particle-hole excitations. The same result for the viscosity Eq.~\eqref{gamma T to 0} can also be obtained by considering the energy dissipation of an infinite mass impurity moving in the bath (see Appendix \ref{Appendix B}), as discussed in Ref.~\cite{astrakharchik2004motion} for the case of interacting Bose gases. Note that, in the latter case (and for any superfluid system), the viscosity vanishes for $T\rightarrow 0$, due to the existence of the critical Landau velocity, which provides a minimal velocity for the impurity to excite the system. Interestingly, Schecter and Kamenev applied the same formalism we adopted in the present work to compute the friction in a weakly interacting Bose gas, and found that it  scales as $\gamma_{BEC}\simeq T^7$~\cite{schecter2014phonon}.

\subsection{Dynamics of two impurities} \label{Subsec: generalized Langevin equation for 2 impurities}

Let us now consider the case of two impurities. The corresponding GLEs read:  
\begin{align}
m_I \ddot{\rg}_1&=-\left(\gamma\dot{\rg}_1+\Gamma_{12}(\rg_1-\rg_2)\dot{\rg}_2 \right)+ \nonumber \\
&+{\bf F}_1(\rg_1-\rg_2)+\mbf{\Psi}_1(\rg_1-\rg_2,t) \label{gle r1} \\ 
m_I\ddot{\rg}_2&=-\left(\Gamma_{21}(\rg_1-\rg_2)\dot{\rg}_1+\gamma\dot{\rg}_2 \right)+ \nonumber \\
&-{\bf F}_1(\rg_1-\rg_2)+\mbf{\Psi}_2(\rg_1-\rg_2,t), \label{gle r2} 
\end{align} 
in strong analogy with the old result for heavy particles in incompressible fluids~\cite{Oppenheim}. It is convenient to rewrite the previous equations in terms of the relative distance  between the impurities, $\mbf{s}=\rg_1-\rg_2$, and  the center of mass $\mbf{r}_{CM}=(\rg_1+\rg_2)/2$:
\begin{align}
m_I\ddot{\mbf{s}}&=- (\gamma-\Gamma_{12}({\bf s}))~ \dot{\mbf{s}}+2{\bf F}_1 (\mbf{s})+\mbf{\eta}_-({\bf s},t) \label{distance gle} \\
m_I\ddot{\mbf{r}}_{CM}&=- (\gamma+\Gamma_{12}({\bf s}))~\dot{\mbf{r}}_{CM}+  \frac{1}{2}\mbf{\eta}_{+}({\bf s},t) \label{c.o.m. gle}, 
\end{align}
where $\mbf{\eta}_+({\bf s},t)$ and $\mbf{\eta}_-({\bf s},t)$ are two Gaussian noises,
\begin{equation}
\eta_\pm({\bf s},t) = \bm{\Psi}_1(\mbf{s},t)\pm\bm{\Psi}_2(\mbf{s},t).
\end{equation}
Using Eqs.~(\ref{real potential position}, \ref{rescaled potential}), the explicit expression for the force and the viscosity matrix can be respectively written as 
\begin{eqnarray}\label{F1}
{\bf F}_1^i (\mbf{s}) &=&\frac{m g^2}{16\pi^4\hbar^2}\frac{s_i}{s^2}\nonumber\\
&&\int_0^{\Lambda}\!\!\!\! dq \;qh(q,s)\int_0^{\infty}\!\!\!\! dk~k f_{FD}(k/2) \log \left\vert \frac{k+q}{k-q} \right\vert \nonumber\\ \label{gradient V} 
\end{eqnarray}
and
\begin{eqnarray}
\Gamma_{12}^{ij}(\mbf{s}) &=& -\frac{m^2 g^2}{4\pi^3 \hbar^3s^2}\int_0^{\infty}\!\!\!\! dq~q\left\lbrace h(q,s)\left(\delta_{ij}-\frac{s_i s_j}{s^2}\right) \right. \nonumber \\
 && \left. -(2h(q,s) + q s \sin (qs))\frac{s_i s_j}{s^2}   \right\rbrace f_{FD}(q/2),\nonumber\\ 
 \label{Gamma12}
\end{eqnarray}
where $h(q,s)=\cos(qs)-\sin(qs)/(qs)$. Note that in Eq.(\ref{F1}) we have introduced a UV momentum cutoff $\Lambda$. To be consistent with the physical interaction characterized by the $s$-wave scattering length $a$, the coupling constant must satisfy~\cite{pethick2008bose}:
\begin{equation}
    4 k_F a = \left( \frac{\hbar^2}{2m_r}\frac{\pi}{g k_F} + \frac{\Lambda}{\pi k_F} \right)^{-1}.
\end{equation}

The presence of a second impurity significantly modifies the stochastic equations of motion. In particular, the relative motion experiences the effect of an external force, which  provides the finite temperature generalization of the RKKY interaction \cite{ruderman1954indirect,kasuya1956theory,yosida1957magnetic}.  In this case,  the friction matrix depends on the relative distance between the pair, distinguishing between friction in the directions collinear and transverse to the relative distance ${\bf s}$. 
In particular, the relative motion becomes underdamped in the limit in which the distance between the impurities is small. 
On the other hand, the center of mass  diffuses according to a simple Brownian motion, with a friction matrix that depends on the relative coordinate only. 

The conservative potential and  the friction matrix elements are plotted in Fig. \ref{Fig: V and Gamma}, which also shows that the temperature dependence is very weak and that  non-collinear friction is small.

 \begin{figure}[t]

\begin{minipage}[t]{0.5\textwidth}
       \includegraphics[keepaspectratio=true,width=8.6cm]{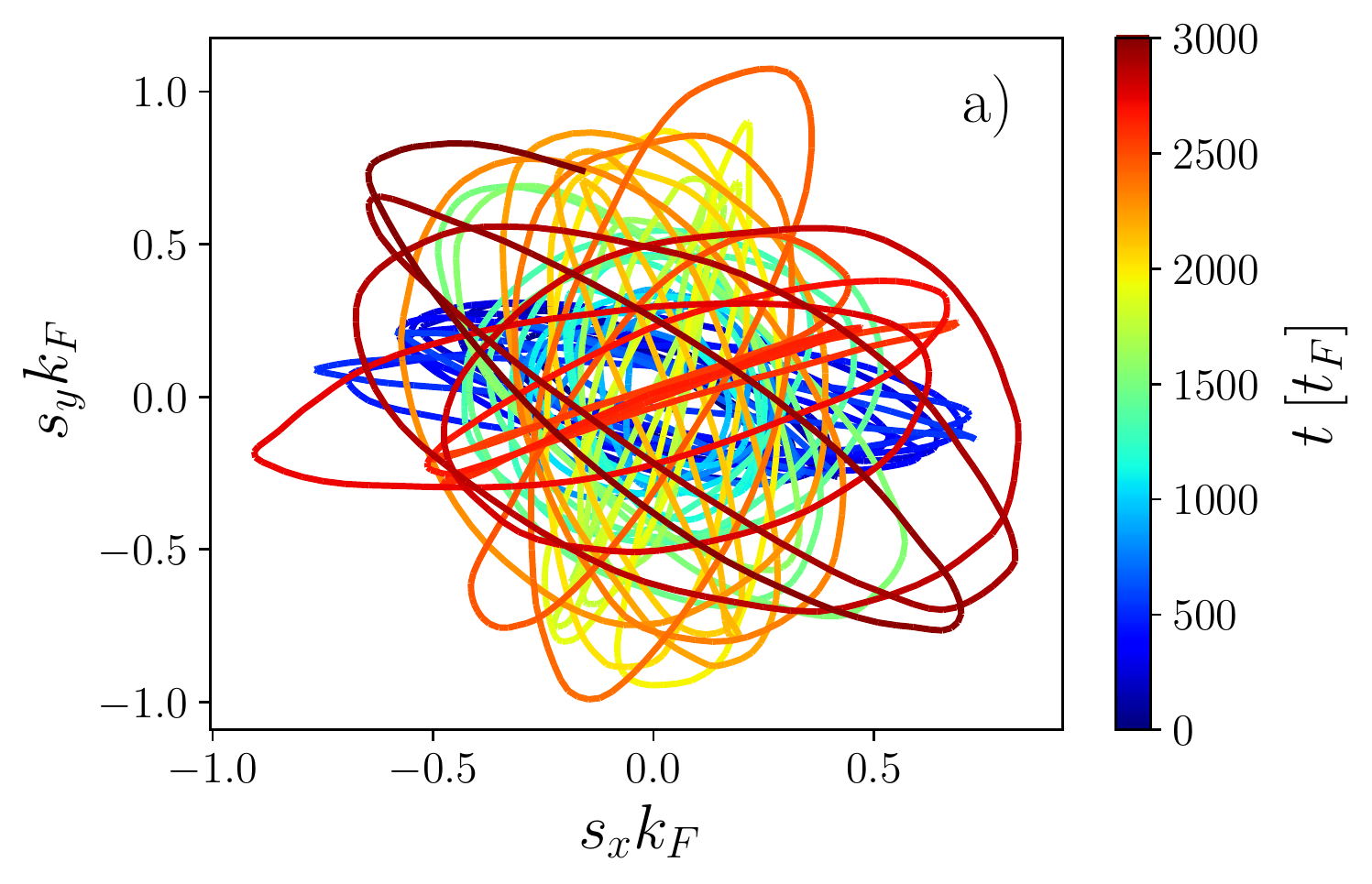}
     \end{minipage}
     \hfill
     \begin{minipage}[t]{0.5\textwidth}
        \includegraphics[keepaspectratio=true,width=8.6cm]{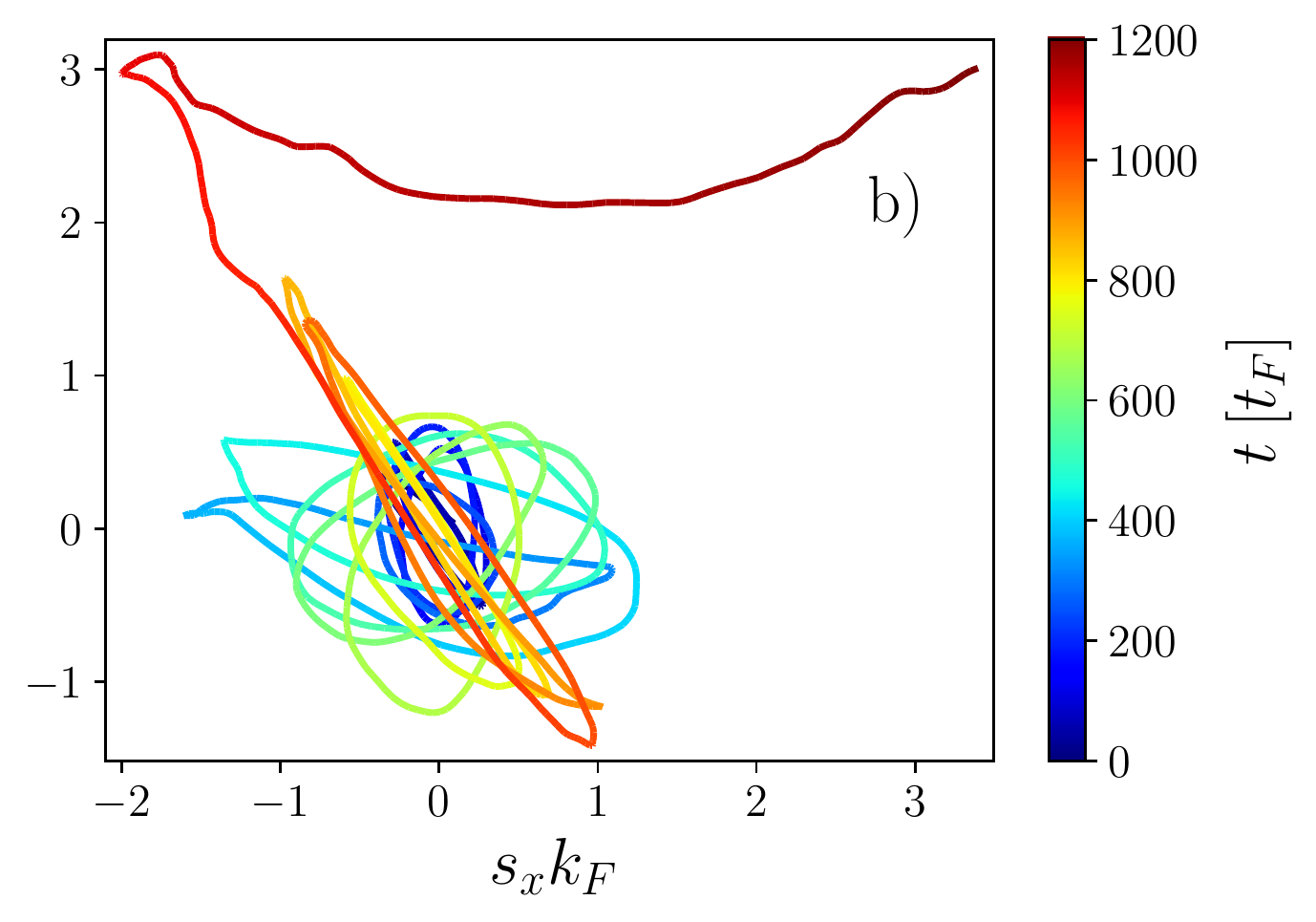}
     \end{minipage}
\caption{\label{Fig: orbit}  
Two representative trajectories obtained by integrating the GLEs starting from a configuration with relative impurity distance $r_ {\textrm b}$, $k_F a =0.19$ and $T/T_F = 0.05$. The components $s_x$, $s_y$ of the distance ${\bf s}$ are shown and the color map labels  time. The motion of the impurities is also not confined on a plane due to thermal fluctuations.  In trajectory reported in panel a), the impurities remain in a bound state throughout the entire simulation time. In panel b) the bound state starts to dissociates for $t \simeq 1000 \, t_F$. In both cases, the relative motion in the bound state becomes quasi one-dimensional, because of angular momentum dissipation induced by the large transverse viscosity. }
\end{figure}

\begin{figure}[t]

\begin{minipage}[t]{0.5\textwidth}
         
        \includegraphics[keepaspectratio=true,width=8.6cm]{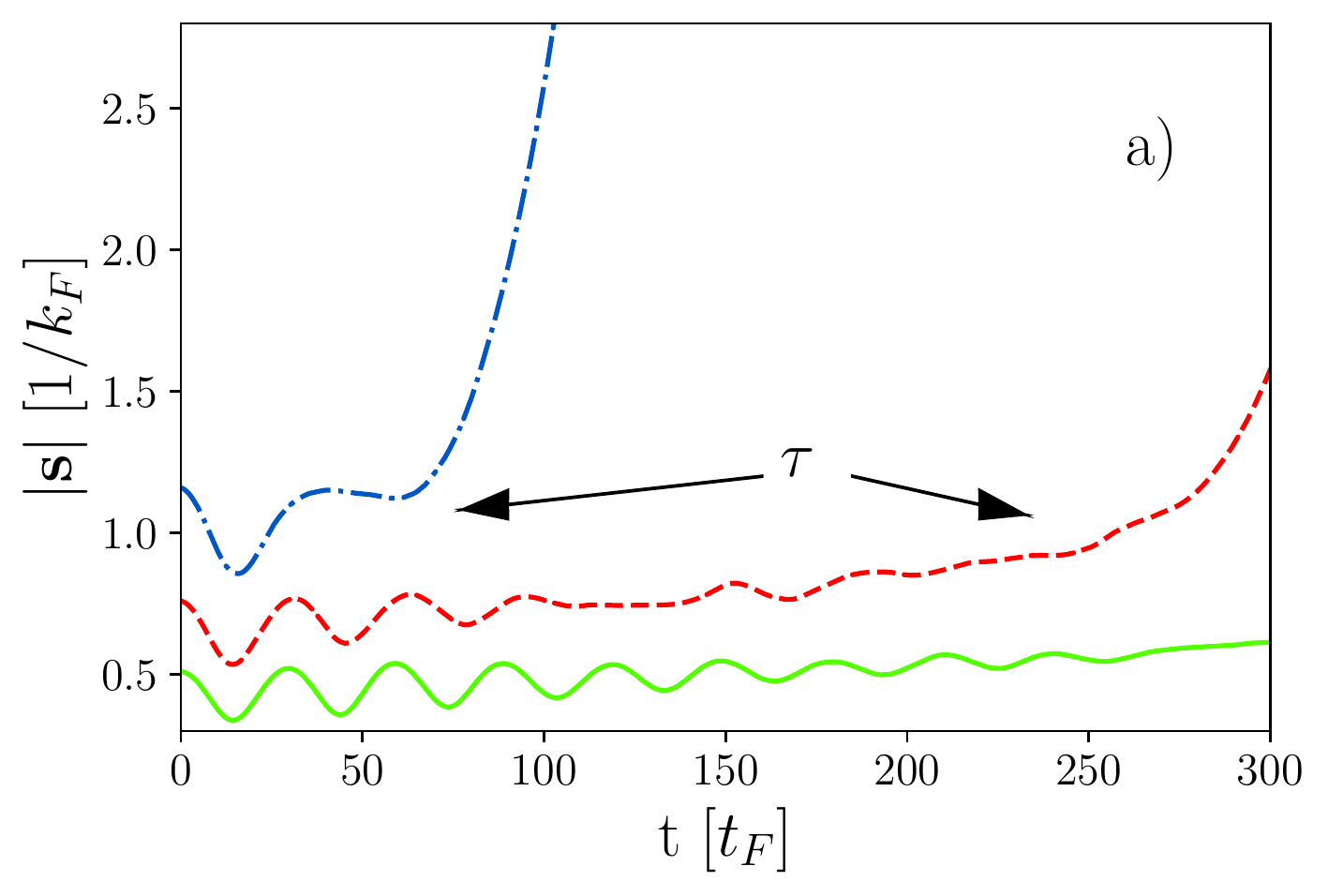}
     \end{minipage}
     \hfill
     \begin{minipage}[t]{0.5\textwidth}
         
        \includegraphics[width=8.6cm]{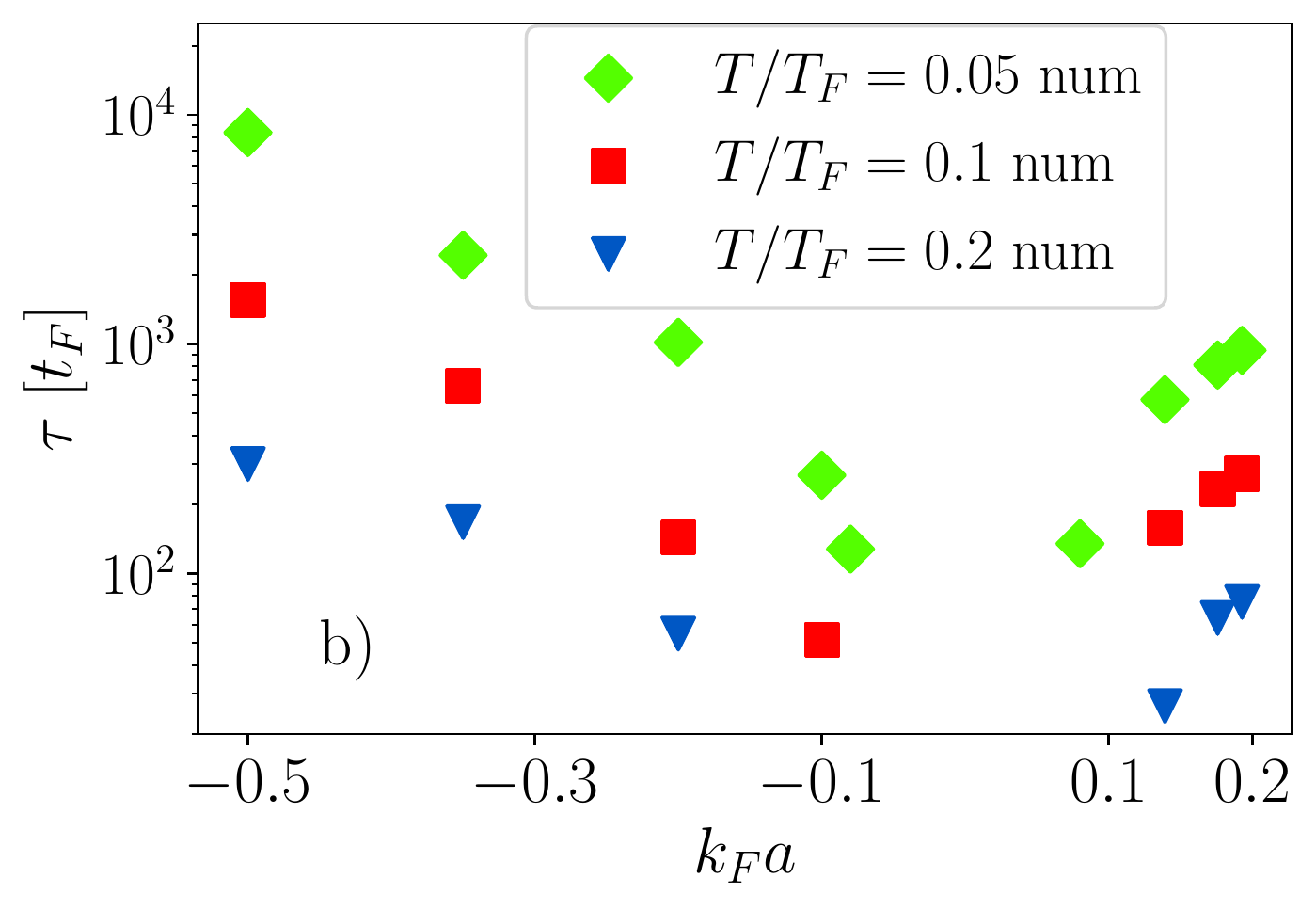}
     \end{minipage}
\caption{\label{Fig: radius} Panel a): time evolution of $s=|\mbf{s}|$ at $k_F a = 0.19$ for different temperatures: full green line is $T/T_F = 0.05$, dashed red line is $T/T_F=0.1$ and blue dotted-dashed line is $T/T_F = 0.2$. For $T/T_F =0.2$ and $T/T_F = 0.1$ also the dissociation of the bound state at time $\tau$ is visible and it is indicated with an arrow. Panel b): Numerically observed life-times $\tau$ (in units of Fermi time $t_F$) at different $T/T_F$ and $k_F a$. }
\end{figure}

\section{Numerical results: bound states dynamics and life-time} \label{Sec: numerical results}

As a case study for the dynamics of two heavy impurities in a free Fermi gas, we focus on the presence of localised (bound) solutions due to the mediated interaction and estimate its life-time under the effect of the stochastic noise. Firstly, we consider the typical value of the bound state size, $r_{\textrm{b}}$: this  can be estimated by matching the average kinetic energy provided by the coupling with the bath with the strength of the mediated interaction:  $(k_F a)^2V(r_ {\textrm b}) \propto 3 k_B T /2$. As expected,  an increase in the scattering length (temperature) leads to a smaller (larger) $r_ {\textrm b}$, as shown in  Fig. \ref{Fig: V and Gamma} a).

We solve Eq. (\ref{generalized langevin equation (complete)}) using a stochastic Verlet algorithm~\cite{ladd2009numerical} and simulate the time evolution of 2 impurities at different temperature, with  scattering length $k_F a $. 
The two impurities start at rest with an initial random position ${\bf s}_0$ subject to the constraint $|{\bf s}_0| = r_b$. 
We average over 1000  independent simulations, with a mass ratio  set to $m_I/m=30$, which is comparable to that of typical experimental setups (for example,  in $^{133}$Cs-$^6$Li mixtures \cite{CsLi} one has $m_I/m\simeq 22$).

In all simulations, we find that in the long-time regime, impurities drift apart and eventually diffuse according to the single-impurity Brownian dynamics described by Eq.(\ref{Lang1}). In Fig. \ref{Fig: orbit}, two representative  trajectories of the short-time regime are shown: in panel a) the impurities remain within a distance comparable to $r_ {\textrm b}$ throughout the entire simulation time, signaling the existence of a bound state. In panel b) the impurities eventually dissociate and begin an independent Brownian diffusion. 

An interesting feature that can be inferred from these trajectories is that the relative motion of the two impurities in the bound state tends to become quasi one-dimensional. This is due to the presence of the transverse component of the friction that leads to the dissipation of the internal orbital angular momentum.

The life-time of the bound state $\tau$ is defined as the average dissociation time. In the low temperature regime (i.e. when dissociation is a thermally activated process), $\tau$ can be calculated using Kramers' theory \cite{hanggi-1990-reaction}:
\begin{equation}
\tau = 2\pi \sqrt{\frac{K}{K_a}}\frac{2 m_I e^{\beta U }}{\sqrt{\gamma^2 + 4K m_I }-\gamma}, \label{lifetime}
\end{equation} 
Here, the viscosity $\gamma$ is estimated from Eq. (\ref{viscT>0}) by taking the limit of vanishing distance,  $K$ and $K_a$ are the curvature of the potential at the top and bottom of the potential energy barrier, and  $U$ is the height of the barrier.

The life-time $\tau$ of the bound state can also be directly inferred from the numerical simulations. Typical evolution of the inter-impurity distance is shown in Fig \ref{Fig: radius} a) for $k_F a=0.19$ and  for $T/T_F=0.2$ (dotted-dashed blue line), $T/T_F=0.1$ (dashed red line) and  for $T/T_F = 0.05$ (green line). For the latter the dissociation occurs at $t>300$~t$_F$. 
In Fig. \ref{Fig: radius} b) the life-time $\tau$ for different scattering lengths is reported and they perfectly agree with the ones predicted by Kramers' Eq. (\ref{lifetime}). This implies that, at these temperatures,  the dissociation of the bound states is a thermally activated event. The range of temperatures we consider is experimentally accessible. In addition, typical Fermi time $t_F = \hbar/\varepsilon_F$ in recent experiments, see e.g.~\cite{scazza2017repulsive}, is of order $10^{-2}$ ms, thus dissociation times between 100 and 1000 $t_F$ should be experimentally detectable.  We stress that an agreement between Kramers' theory predictions and experimental dissociation times would represent a validation of the classical approach developed in this work.

\section{Conclusions} \label{Sec: conclusions}
In  a fermionic bath, the stochastic dynamics of  impurities is strongly influenced by the effective interaction and friction induced by the coupling to the medium. Under a well-controlled chain of approximations, the bath degrees of freedom can be traced out and the impurities' dynamics can be described through an effective stochastic dynamics. 
In this  scheme, the impurities obey classical GLEs, and the quantum nature of the system is encoded only in the induced force and viscosity terms.

In this work, we focused on the dynamics of a system consisting of two impurities. 
We found that, in the short time regime, the interplay between  induced interaction and  thermal fluctuations leads to the formation of a bound state characterized by a radius $r_ {\textrm b}$ and a life-time~$\tau$.

Two experimental realizations of mixtures of Bose-Einstein condensates and Fermi gas have been achieved so far \cite{DeSalvo2019, DavidsonMediated2020}. However, the density of  heavy impurities in these systems is relatively high, so that a description in terms of  heavy particles  independently diffusing in the medium may not be accurate. An important question to address is whether it is feasible to experimentally probe systems with lower impurity densities, using the existing technology.  

Although more demanding, cold gases could also be the proper platform to obtain a direct experimental evidence of a non-collinear friction for the impurities.

We note that the same approach adopted in the present work was applied by some of us to investigate the dynamics of heavy quarks diffusing in a ultra-relativistic quark-gluon plasma  Ref.~\cite{blaizot2016heavy}. That analysis was based on an effective finite temperature Abelian gauge theory, to describe the dynamics in the deconfined plasma. In that approach, heavy quarks and anti-quarks played the role of two distinct types of impurities, while light quarks and anti-quarks formed  the thermal bath. All quarks in the systems were coupled via a Debye-screened Coulomb-type interaction.  As a consequence of these features, the sign of $\Gamma_{12}$ was found to be different from that of the present Fermi system. Namely, the center of mass motion experiences a very reduced effective friction, while the relative internal motion of the quark-antiquark pair is overdamped.
As an outlook, it could be interesting to device a cold atom system that can mimic such a model. This may be done by properly selecting two different hyperfine levels or two different atomic species that couple with opposite sign to the particles in the Fermi bath.  The extension of the present simulation strategy to a superfluid fermionic bath and to many-body systems of impurities would also be extremely valuable to understand the properties of the outer layers of neutron stars, such as entrainment effects caused by the presence of the medium (see e.g.~\cite{CHAMEL2005109}) and modifications to transport properties of the crust like the thermal conductivity~\cite{Horowitz2009,Roggero2016} and the neutrino opacity~\cite{Horowitz2004,horowitz2016nuclear,Roggero2018}. 
  
\section*{Acknowledgements}
We thank J. P. Blaizot for useful discussions. Financial support from the Italian MIUR under the PRIN2017 project CEnTraL (Protocol
Number 20172H2SC4), from the Provincia Autonoma di Trento and from Q@TN, the joint lab between University of Trento, FBK- Fondazione Bruno Kessler, INFN- National Institute for Nuclear Physics and CNR- National Research Council is acknowledged.

\appendix
\section{Complex potential}\label{Appendix A}

In Sec.~\ref{Sec: model and theory} we introduced the matrix of polarization propagators $\Delta_{ab}$. Here we demonstrate that with a bath in thermal equilibrium all of these functions are related and we only need one, $\Delta^R$, to derive the complex potential needed for the dynamics. 

In position space, $\Delta^R$ is defined as
\begin{equation}
\Delta^R (\xy) = \Delta^F (\xy)-\Delta^< (\xy). \label{app eq: delta r}
\end{equation} 
The procedure to derive Eqs.~(\ref{real potential}, \ref{imaginary potential}) of the main text is modelled on Ref.~\citep{blaizot2016heavy}. Complex potential in small frequency approximation is
\begin{eqnarray}
i\mathcal{V}(\mbf{q}) = \lim_{\omega \to 0}(\Delta^R(\omega, \mbf{q})+\Delta^<(\omega, \mbf{q})) = \nonumber \\
=\lim_{\omega \to 0}(\re \Delta^R(\omega, \mbf{q}) + i \im \Delta^R (\omega, \mbf{q}) + \Delta^< (\omega, \mbf{q})). \nonumber \\
\label{app eq: comp pot 1}
\end{eqnarray}
The real part of $\Delta^R$ is related to the spectral density $\sigma$. Indeed, we have that $2\re \Delta^R(\omega, \mbf{q})=\sigma(\omega, \mbf{q})$. In small frequency approximation therefore
\begin{eqnarray}
\re \Delta^R (\omega, \mbf{q}) = A^R(\mbf{q})+\omega B^R(\mbf{q})+o(\omega^2) .
\end{eqnarray}
Spectral density is odd in $\omega$, i.e. $\sigma(-\omega, \mbf{q})=-\sigma (\omega, \mbf{q})$. Therefore $A(q)=0$ and
\begin{equation}
\re \Delta^R (\omega, \mbf{q}) = \omega B^R(\mbf{q})+o(\omega^2)=\frac{1}{2}\sigma (\omega , \mbf{q}) .
\end{equation}
Exploiting the fluctuation-dissipation relation (FDR), that is valid for a bath at equilibrium, we have that
\begin{equation}
\Delta^< = \frac{2}{e^{\beta \omega}-1}\re \Delta^R (\omega , \mbf{q}). \label{app eq: delta<}
\end{equation}
In this last equality an algebraic relation between $\Delta^<$ and $\Delta^R$ is established. Thanks to this, we will be able to write ${\cal V}$ only in terms of $\Delta^R$.  

The limit $\omega \to 0$ of Eq.~(\ref{app eq: delta<}) is
\begin{equation}
\lim_{\omega \to 0}\Delta^<(\omega , \mbf{q})=\frac{2}{\beta}B^R(\mbf{q}).
\end{equation}
We now perform the limit in Eq.~(\ref{app eq: comp pot 1}) and we obtain
\begin{eqnarray}
\mathcal{V}(\mbf{q})=\im \Delta^R(\omega = 0, \mbf{q})-i \frac{2}{\beta}B^R(\mbf{q}) .
\end{eqnarray}
In terms of $D^R = -i \Delta^R$ the real and imaginary part of the complex potential now are
\begin{eqnarray}
V(\mbf{q}) = \re D^R(\omega =0,\mbf{q}) , \\
W(\mbf{q}) = \frac{2}{\beta}\lim_{\omega \to 0} \frac{\im D^R(\omega,\mbf{q})}{\omega} .
\end{eqnarray}
The full expressions of $\re D^R$ and $\im D^R$ in momentum space are
\begin{align}
&\re D(\omega, \mbf{q}) = -\frac{m}{2\pi^2\hbar^2}\int dk f_{FD}(k,T) \frac{k}{2q} \times \nonumber \\
 &\times\left( \log \left\vert\frac{k/k_F-\nu_-}{k/k_F+\nu_-}  \right\vert - \log \left\vert\frac{k/k_F-\nu_+}{k/k_F+\nu_+}  \right\vert \right), \label{app: real D R}\\
 &\im D(\omega, \mbf{q}) = -\frac{m k_F}{2\pi\hbar^2}\left[\frac{\omega}{v_F q}+\frac{1}{\beta v_F q}\log \left(\frac{1+e^{\beta(\nu_-^2\varepsilon_F-\mu)}}{1+e^{\beta(\nu_+^2\varepsilon_F-\mu)}} \right) \right], \label{app: imag D R}
\end{align}
where $\beta = 1/k_B T$, $\nu_{\pm}=\omega/q v_F \pm q/2k_F$ and $v_f = k_F/m$. 

\section{Zero temperature friction}\label{Appendix B}
As shown in Ref.~\cite{astrakharchik2004motion}, friction can be understood also in terms of energy dissipated by an impurity moving at velocity $V$, $\dot{E} = -F_V V$, with $F_V$ the velocity dependent drag force. Following the convention of \cite{pitaevskii2016bose}, Sec. 7 we have that the energy dissipated per unit time and unit particle when a contact interaction of strength $g$ is considered is
\begin{align}
    \dot{E} &= -\int_{-\infty}^\infty \frac{d{\bf k}}{(2\pi)^3} \!\! \int_{-\infty}^\infty \frac{d \omega}{2\pi} 2\pi S(\omega, {\bf k})\frac{n}{2N}\omega 2\pi g^2\delta(\omega - k_z V)= \nonumber \\
    &=-\frac{n g^2}{2N}\frac{1}{(2\pi)^2}\int_{-\infty}^\infty d{\bf k}~S(k_z V,{\bf} k) k_z V = -F_V V.
    \label{E dot}
\end{align}
Now we focus on the drag force $F_V$
\begin{align}
    F_V &= \frac{n g^2}{8N \pi^2}\int_{-\infty}^\infty \! d{\bf k}S(k_z V,{\bf k})k_z = \nonumber \\
    &= \frac{n g^2}{4N \pi}\iint_{-\infty}^{\infty}\!\! dk_\perp dk_z k_\perp k_z S(k_z V,\sqrt{k_z^2+k_\perp^2}) \label{F_V 1}.
\end{align}
In order to perform the integration in \eqref{F_V 1}, we use the expression of the dynamical structure factor $S(k_z V,\sqrt{k_z^2+k_\perp^2})$ given in Ref.~\cite{nozieres2018theory}, Sec. 2. This expression is 
\begin{equation}
    S(\omega, {\bf k}) = \frac{\nu (0)}{2}\frac{\omega}{k v_F} \quad \text{if} \quad 0 \leq \omega \leq kv_F - \frac{k^2}{2m},
\end{equation}
that in the small velocity limit gives the conditions $0\leq k_z \leq 2m v_F$ and $0 \leq k_\perp \leq \sqrt{(2m v_F)^2 - k_z^2}$. Performing the integration we obtain
\begin{multline}
    \frac{\nu (0) V}{2v_F}\int_0^{2m v_F}\!\!dk_z~k_z^2 \int_0^{\sqrt{(2m v_F)^2-k_z^2}}\!\!dk_\perp \frac{k_\perp}{\sqrt{k_z^2+k_\perp^2}}= \\
    =\frac{8}{3}\nu(0)mV k_F^3 .
\end{multline}
Finally, for $F_V$ we obtain
\begin{align}
    F_V &= \frac{n g^2}{4N \pi} \frac{8}{3}\nu(0)mV k_F^3 = \frac{3mN}{k_F^2}\frac{k_F^3}{6\pi^2}\frac{g^2}{N\pi}\frac{2}{3}m k_F^3 V = \nonumber \\
    &=\frac{m^2 k_F^4}{3\pi^3}g^2 V = \frac{4k_F^2}{3\pi}\left(k_F a \frac{m}{m_r} \right)^2 V = \gamma_{T=0} V. \label{F_V final}
\end{align}
In this derivation, we used $\nu(0) = 3mN/k_F^2$, see \cite{nozieres2018theory}, and $n=k_F^3/6\pi^2$. 
Now, comparing \eqref{F_V final} with \eqref{gamma T to 0} we see that we recovered the same result for the friction coefficient at $T=0$ (the missing $\hbar$ factor is due to the fact that in this Appendix we set $\hbar =1$). 

This connection between the statistical structure factor $S(\omega, {\bf k})$ gives also a useful insight on why $\gamma$ vanishes for a Bose gas or generally for a phononic spectrum at $T=0$. 
The dynamical structure factor in presence of  single low energy phonon mode reads
$S(\omega, {\bf q}) = S_{\bf k}\delta (\omega - c|\bf k|)$, where $c$ is the speed of sound.
Therefore the drag force vanishes for any impurity speed $V<c$ (obviously in agreement with the Landau criterion for superfludity).

On the other hand, having the fermions a continuum of particle-hole excitations at low energy, a moving object will release energy to the bath at whatever speed $V$ it moves.

\bibliography{Bibliography_QBM} 
\end{document}